\documentclass[final]{aa}

\title{Survey of the  ISM in Early-Type Galaxies. IV. The Hot Dust Component
\thanks{ Based on observations with ISO, an ESA project with instruments 
founded by ESA member states (especially the PI countries: France, Germany, 
the Netherlands and the United Kingdom) and with participation of 
ISAS and NASA.}}

\author{F. Ferrari\inst{1} \and M.G. Pastoriza\inst{1} \and F.D. Macchetto\inst{2}$^,$\inst{3}
  \and C. Bonatto\inst{1} \and  N. Panagia\inst{2} \and W. B. Sparks\inst{2} }     

\offprints{Fabricio Ferrari, \email{ferrari@if.ufrgs.br}}

\institute{ Instituto de Fisica, UFRGS, Porto Alegre, Brazil \and 
Space Telescope Science Institute \and 
On assigment from the Space Science, Department of ESA }

\date{Received date, Accepted 2002 April 16}

\titlerunning{ISM in Early-Type Galaxies: The Hot Dust Component}
\authorrunning{Ferrari et al. }

\usepackage{graphicx}
\usepackage{pifont}
\usepackage{array}
\usepackage{longtable}

\abstract{ We present mid-IR photometric properties for a sample of 28
  early-type galaxies observed at 6.75, 9.63 and 15 $\mu{} $m with the ISOCAM
  instrument on board the ISO satellite.  We find total mid-IR luminosities in
  the range $3-48 \times{} 10^{8}$ $ L_{\odot}$. The spectral energy distribution (SED)
  of the galaxies were derived using the mid-IR data together with previously
  published UV, optical and near-IR data.  These SEDs clearly show a mid-IR
  emission coming from dust heated at $T \simeq 260$ K.  Dust grains properties
  are inferred from the mid-IR colors. The masses of the hot dust component are
  in the range $10-400$ $M_{\odot}$.  The relationship between the masses
  derived from mid-IR observations and those derived from visual extinction are
  discussed. The possible common heating source for the gas and dust is
  investigated through the correlations between ${\rm H}\alpha$ and mid-IR
  luminosities.
\keywords{ galaxies: early-type, interestellar medium}
}

\begin{document}
\maketitle

\section{Introduction}
\label{sec:intro}

The presence and properties of the interstellar medium (ISM) in early-type
galaxies have been the subject of intense work in the last decades.  In
particular, dust lanes and patches have been detected in a large fraction of
such galaxies, typically in about 75\% of the objects. The morphology of the
dust distribution in these galaxies follows very closely that of the ionized gas
(Ferrari et al. 1999, Goudfrooij et al. 1994a) and typical dust masses in the
range $10^3-10^5$ $M_{\odot}$ have been found (Ferrari et al. 1999; Goudfrooij
et al.  1994b).  These results were confirmed also by optical HST data, which
showed that 78\% of the early-type galaxies contain nuclear dust (Van Dokkum \&
Franx 1995).  However, dust masses derived from the IRAS flux densities (Knapp
et al. 1989) are found to be roughly an order of magnitude higher than those
derived from the optical extinction values. This is in strong contrast with what
is found in spiral galaxies where dust masses derived from the optical
extinction and IRAS luminosities are in agreement (Goudfrooij et al. 1995,
Merluzzi 1998). To account for this difference it was argued that most of the
dust in elliptical galaxies exists as a diffuse component essentially
undetectable at optical wavelengths. This diffuse dust component produces a
radial color gradient that adds to the gradient produced by metallicity and
stellar population age, as shown by Goudfrooij \& Jong (1995).

Knowledge of the ISM dust emission properties in galaxies has greatly increased
over the last years due to the ISO satellite which has provided high quality
data in the spectral range 3 - 200 $\mu{} $m.  The IR emission of the interstellar
medium of galaxies is mainly due to dust heated by the interstellar radiation
field and a large fraction of this emission is observed in the near and mid-IR.
The emission in this region has been explained under the hypothesis of a dust
population formed by very small ($a < 0.02\;\mu{} $m) grains with a
fluctuating temperature, due to the absorption of individual UV photons.
Emission in the wavelength range $3 < \lambda < 15 \;\mu{} $m is also produced by
Unidentified Infrared Emission Bands (UIBs) associated with the Polycyclic
Aromatic Hydrocarbons (PAHs) (Desert et al.1990).  The mid-IR emission of a
selected sample of Virgo cluster early-type galaxies has been studied by Boselli
et al. (1998), who concluded that the emission up to 15 $\mu{} $m is dominated by
the Rayleigh-Jeans tail of the old stellar population.  A survey of a larger
number of E and SO galaxies using broad band filter ISOCAM imaging covering the
range 4.5 to 18 $\mu{} $m shows that dust present in the form of small hot grains
and/or UIB also contributes to the mid-IR emission in these galaxies (Madden et
al.1999).

The origin of the gas and dust in early-type galaxies is still a highly
controversial subject. Alternative scenarios include a normal quiescent
component, accretion by merger events and cooling flows (Sparks et al. 1989, Kim
1989, Fabian 1994). In order to contribute to the investigation of the ISM
origin in early-type galaxies, we have carried out an extensive program of
imaging and spectroscopy of a large sample of bright E and S0 galaxies. The
ionized gas properties in these galaxies are discussed in Macchetto et al. (1996,
hereafter Paper I).  The morphology of the dust distribution and the optical
absorption $A_V$ together with assumptions on dust grain composition allowed us
to estimate the dust masses for the galaxy sample. These results were presented
in Ferrari et al. (1999, hereafter Paper II). Gas and star kinematics for 12 of
these galaxies are discussed in Caon et al. (2000).

In the present paper we discuss the properties of the mid-IR emission in 28
early-type galaxies observed with the ISOCAM instrument on board the ISO
Satellite for which we have obtained broad band imaging, covering the range 5 to
18 $\mu{}$m.  For each galaxy we have estimated the total mid-IR luminosity, mid IR
colors and dust properties.  We have compared these results with those
inferred from the optical absorption and explored the correlation between the
mid-IR dust luminosity with the ionized gas luminosity given in Paper I and dust
masses derived from optical data from Paper II.
 
This paper is structured as follows: in Section \ref{sec:sample} we present the
galaxy sample; in Section \ref{sec:obs} we discuss the observations and data
reduction.  The Spectral Energy Distributions of the target galaxies are derived
in Section \ref{sec:sed}.  The luminosity distribution at the three observed
frequencies is presented in section \ref{sec:lumin}, and mid-IR colors in
Section \ref{sec:midir}. Dust mass estimates from the mid-IR observations are
presented in Section \ref{sec:mass}. Finally, the conclusions are given in
Section \ref{sec:conclusions}.

\section{The sample}
\label{sec:sample}

The objects studied in this paper, extensively described in Papers I and II, are
luminous galaxies ($B_T < 13$) with morphological types E and SO selected from
``The Reference Catalog of Bright Galaxies'' (RC3) to provide a fairly large
interval in optical luminosity including both radio-loud and radio-quiet
galaxies, X-ray emitters and non-emitters.  A large fraction of the galaxies in
this sample (72\%) contains ionized gas (Paper I) and considerable amounts of
dust (Paper II). This sample is not complete in any sense, rather it includes
rich ISM galaxies.  14 galaxies have IRAS upper limits only; 8 were detected 
in 2 IRAS bands and have 2 upper limits and only 3 have reliable IRAS data.
Our sample galaxies are listed in Table \ref{table:nomes} along
with their coordinates, morphological type, magnitudes, distances and stellar
population groups (discussed later).

\begin{footnotesize}
\begin{table*}
\begin{center}
\begin{tabular}{l  >{$}l<{$} >{$}l<{$} l l l l } \hline
Name &\alpha {\rm(2000)} &\delta {\rm (2000)}&Type &$B^0_T$ & D (Mpc)& SED \\ \hline  
NGC 720   &  01\;52\;57.5  &  -13\;30\;20  &  E5     &  11.16   & 25.4 & --  \\
NGC 741   &  01\;56\;21.0  &  +05\;37\;44  &  E0     &  12.2    & 84.0 & --  \\
NGC 1453  &  03\;46\;27.2  &  -03\;58\;09  &  E2     &  12.58   & 59.1 & K7  \\
NGC 3258  &  10\;28\;54.1  &  -35\;36\;22  &  E1     &  12.22   & 42.9 & --  \\
NGC 4374  &  12\;25\;03.7  &  +12\;53\;13  &  E1     &  10.01   & 15.8 & K4  \\ 
NGC 4472  &  12\;29\;46.8  &  +08\;00\;01  &  E2/S0  &  9.33    & 18.0 & K4  \\
NGC 4636  &  12\;42\;50.0  &  +02\;41\;17  &  E/S0   &  10.43   & 18.0 & --  \\ 
NGC 4783  &  12\;54\;36.3  &  -12\;33\;30  &  E0pec  &  12.80   & 71.2 & --  \\
NGC 4936  &  13\;04\;16.4  &  -30\;31\;29  &  E0     &  12.28   & 50.4 & K7  \\
NGC 5044  &  13\;15\;24.0  &  -16\;23\;06  &  E0     &  11.87   & 44.2 & K7  \\
NGC 5084  &  13\;20\;16.7  &  -21\;49\;39  &  S0     &  11.28   & 26.6 & M0  \\
NGC 5813  &  15\;01\;11.2  &  +01\;42\;08  &  E1     &  11.42   & 29.5 & K4  \\
NGC 5831  &  15\;04\;07.2  &  +01\;13\;15  &  E4     &  12.31   & 29.5 & M0  \\ 
NGC 5903  &  15\;18\;36.3  &  -24\;04\;06  &  E3/S0   &  11.74  & 37.5 & K7  \\ 
NGC 6407  &  17\;44\;57.7  &  -60\;44\;22  &  E       &  12.43  & 67.9 & K4  \\ 
NGC 6721  &  19\;00\;50.5  &  -57\;45\;28  &  E1      &  12.68  & 64.9 & M0  \\ 
NGC 6758  &  19\;13\;52.3  &  -56\;18\;33  &  E2      &  12.31  & 49.7 & K7  \\ 
NGC 6776  &  19\;25\;19.4  &  -63\;51\;41  &  E1pec   &  12.71  & 79.9 & M0  \\ 
NGC 6851  &  20\;03\;33.6  &  -48\;17\;02  &  E4      &  12.51  & 41.1 & K4  \\ 
NGC 6868  &  20\;09\;53.8  &  -48\;22\;45  &  E3/S0   &  11.49  & 41.1 & K7  \\ 
NGC 6876  &  20\;18\;20.3  &  -70\;51\;28  &  E3      &  11.83  & 57.4 & --  \\ 
NGC 7041  &  21\;16\;32.6  &  -48\;21\;51  &  S0/E7   &  12.19  & 27.7 & K7  \\ 
NGC 7562  &  23\;15\;57.7  &  +06\;41\;15  &  E2-3    &  12.38  & 51.9 & M0  \\ 
NGC 7619  &  23\;20\;14.7  &  +08\;12\;23  &  E       &  11.93  & 51.9 & M0  \\ 
NGC 7626  &  23\;00\;42.3  &  +08\;13\;02  &  Epec    &  12.06  & 51.9 & --  \\ 
NGC 7796  &  23\;58\;59.7  &  -55\;27\;23  &  E+      &  12.39  & 47.7 & K4  \\ 
IC 4889  &  19\;45\;15.8  &  -54\;20\;37   &  E5      &  12.06  & 36.8 & --  \\ 
IC 5105  &  21\;24\;22.4  &  -40\;32\;06   &  E+      &  12.62  & 79.1 & M0  \\ 
\hline
\end{tabular}
\caption{Names, coordinates, morphological types,  integrated total blue
  magnitudes and distances ($H_0=65\;{\rm km\,s}^{-1}\,{\rm Mpc}^{-1}$) for the
  galaxies in the sample. SED describes the color of the dominant stellar
  contribution (see text); -- means galaxies not classified in any group.}
\label{table:nomes}
\end{center}
\end{table*}
\end{footnotesize}

Distances have been derived with the ``220 model'' for the Virgo Infall of
Kraan-Korteweg (1986) assuming the Virgo Cluster at a distance of 21.3 Mpc.  For
galaxies not present in the Kraan-Korteweg's list, distances were derived from
their redshift corrected to the reference frame defined by the cosmic microwave
radiation (from RC3). All distances were then rescaled according to 
the value of ${\rm H}_0=
65\,{\rm Km\, s}^{-1}{\rm Mpc}^{-1}$ (Macchetto et al.  1999).

\section{Observations and data reductions}
\label{sec:obs}

The observations were carried out with the $32\times{} 32$ element infrared camera
(ISOCAM) on board the ISO satellite. ISOCAM was used to perform a raster image
covering the central body of the targets.  The raster was built by combining the
individual frames (usually more than 100 in each target) slightly shifted along
the observation. The final raster has $36\times{} 36 $ pixels, or $108^{\prime\prime} \times{}
108^{\prime\prime}$, since the pixel field of view (PFOV) is $3^{\prime\prime}$.
The single integration time was 7 sec in most cases.  All galaxies were observed
in 3 filters LW2 ($\lambda_c=6.75\,\mu{} $m , $\Delta\lambda=2.5\,\mu{} $m), LW3
($\lambda_c=15.0\,\mu{} $m, $\Delta\lambda=5\, \mu{} $m) and LW7 ($\lambda_c=9.63\,\mu{} $m,
$\Delta\lambda=2\,\mu{} $m). The ISOCAM data presented in this paper were analyzed
using CIA\footnote{CIA is a joint development by the ESA Astrophysics Division
  and the ISOCAM consortium.  The ISOCAM consortium is led by the ISOCAM PI, C.
  Cesarsky.}: the images were dark-subtracted, flat-fielded, deglitched,
corrected for transient effects in the detector and combined to produce a final
raster image.  At first aproach, we used the mid-IR sky model by Kelsall et al.
(1998), however the resulting fluxes were too large compared with 12 $\mu{}$m IRAS
fluxes.  Consequently, the sky background was measured in the galaxy frames.
In some cases (NGC 720, NGC 4374, NGC4472, NGC 4636) the field of view is
smaller than the size of the galaxy, which could lead to overestimated sky
values and consequently underestimated galaxy fluxes for these galaxies.

\begin{figure}
  \centering \resizebox{\hsize}{!}{\includegraphics{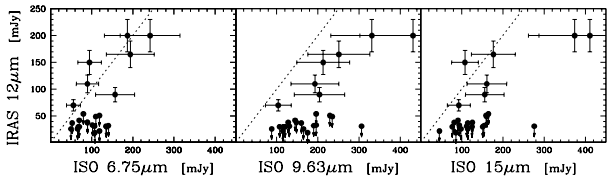}}
\caption{Relation between ISO and IRAS data for the galaxy sample. The straight
  line represents the one-one correlation.}
\label{fig:isoiras}
\end{figure}

The total fluxes were integrated in the raster images after removing the
background and then converted to mJy using the conversion factor in the CIA
packages.  The absolute fluxes are estimated to be accurate to 30\% due to ISO
calibration errors (Cesarsky and Blommaert 2000). It should be mentioned that
the ISOCAM fluxes (\mbox{$50\leq F \leq 374 $} mJy) correspond to the range
where the camera is quite sensitive and accurate. ISO total integrated fluxes
inside the field of view as well as IRAS data are listed in
Table~\ref{table:fluxos}. To test the reliability of ISOCAM data we made a
comparison with IRAS $12 \mu{}$m data (Fig.~\ref{fig:isoiras}). However, 12$\mu{}$m
fluxes fall in a range corresponding to IRAS upper limit detection, consequently
these values are not so reliable as ISOCAM fluxes.  If we do not consider IRAS
$12 \mu{}$m fluxes below $\sim 60$ mJy, which are below the instrument detection
limit, there is a good agreement between ISO $6.75 \mu{}$m and IRAS $12 \mu{}$m fluxes.
The same for ISO $15 \mu{}$m and IRAS $12 \mu{}$m fluxes, except for two galaxies (NGC
4472 and NGC 4636) which are active galaxies. It can also be seen that ISO $9.63
\mu{}$m fluxes are systematically larger than IRAS $12 \mu{}$m fluxes and this may be
due to the presence of some silicate bands that lie in the $9.63 \mu{}$m filter
spectral region. Furthermore, we should not expect ISOCAM and 12$\mu{}$m IRAS fluxes
to agree better than for a factor of 2, because of the differences in the field
of view, filters sensitivity and bandpasses.

\begin{figure}
  \centering \resizebox{0.6\hsize}{!}{\includegraphics{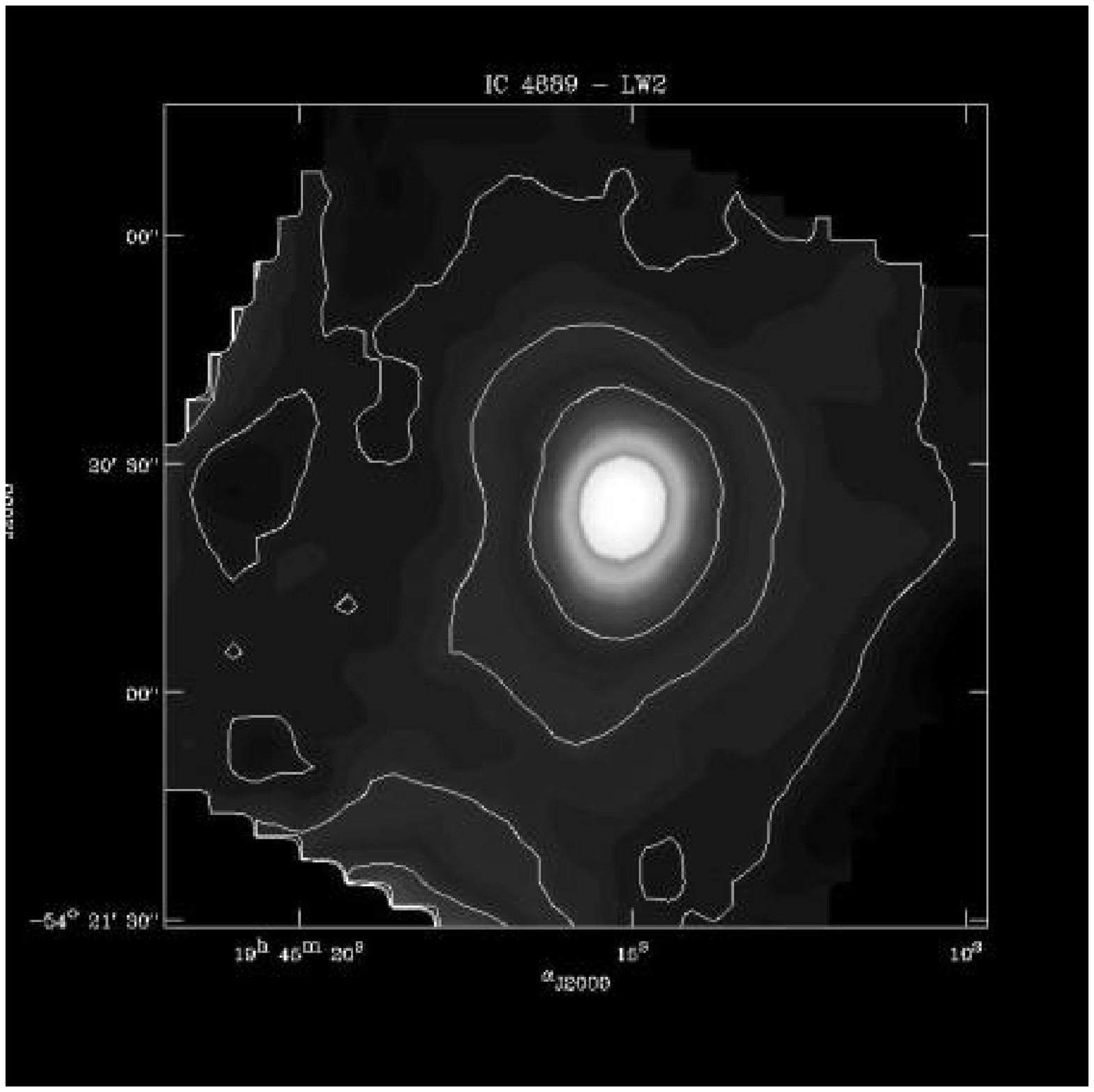}}
\caption{LW2 (6.75 $\mu{}   $m) image of the IC 4889 galaxy.}
\label{fig:ic4889lw2}
\end{figure}

\begin{figure}
  \centering \resizebox{0.6\hsize}{!}{\includegraphics{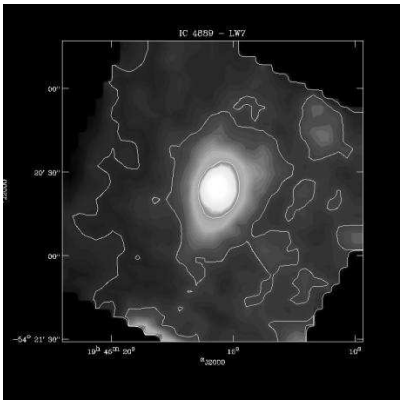}}
\caption{LW7 (9.63 $\mu{}   $m) image of the IC 4889 galaxy.}
\label{fig:ic4889lw7}
\end{figure}

\begin{figure}
  \centering \resizebox{0.6\hsize}{!}{\includegraphics{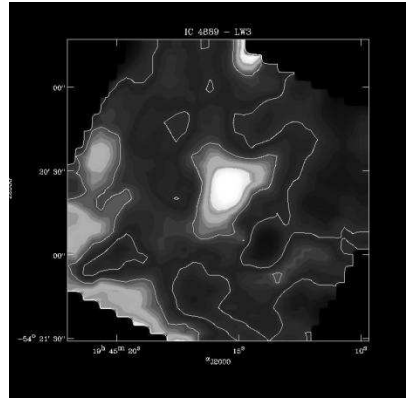}}
\caption{LW3 (15 $\mu{}   $m) image of the IC 4889 galaxy.}
\label{fig:ic4889lw3}
\end{figure}

For most of the observed galaxies, the mid-IR emission is extended, although its
distribution varies among the filters, probably due to the relative contribution
of stars and dust in each filter.  Figures \ref{fig:ic4889lw2},
\ref{fig:ic4889lw7} and \ref{fig:ic4889lw3} present the isophotes of IC 4889, a
typical example of our sample.  The isophotes at 6.75 $\mu{} $m are circular,
corresponding to the dominant contribution of the galaxy stellar bulge, while at
longer wavelengths, where the dust emission begins to dominate, the isophotes
becomes more irregular. A general feature of isophote maps is a concentration of
dust around the optical nucleus of each galaxy.

\begin{footnotesize}
\begin{table}
\begin{center}
\begin{tabular}{l  >{$}r<{$} >{$}r<{$} >{$}r<{$}  >{$}r<{$}   } \hline
          & \multicolumn{4}{c}{Total Fluxes } \\
Galaxy    & 6.75 \mu{}   m   & 9.63 \mu{}   m  & 15 \mu{}   m     & {\rm IRAS} 12 \mu{}m\\ 
          & {\rm\;(mJy)} & {\rm\;(mJy)}& {\rm\;(mJy)} & {\rm\;(mJy)} \\ \hline  

NGC 720     &     157     &        204     &     156      &  90  \\ 
NGC 741     &      65     &        209     &     61       &  --  \\ 
NGC 1453    &      90     &        193     &    161       &  110 \\ 
NGC 3258    &      53     &        160     &    115       &  <37 \\ 
NGC 4374    &     194     &        251     &    177       &  165 \\  
NGC 4472    &     242     &        331     &    374       &  200 \\ 
NGC 4636    &     187     &        431     &    411       &  200 \\ 
NGC 4783    &      70     &        145     &     91       &  <42 \\ 
NGC 4936    &     119     &        115     &     46       &  <22 \\  
NGC 5044    &      95     &        213     &    108       &  150 \\ 
NGC 5084    &     119     &        229     &    159       &  <51 \\  
NGC 5813    &     136     &        116     &    151       &  <30 \\ 
NGC 5831    &      65     &        222     &    152       &  --  \\ 
NGC 5903    &      80     &        196     &    165       &  <54 \\ 
NGC 6407    &      90     &        128     &     82       &  <38 \\ 
NGC 6721    &      50     &        165     &    115       &  <26 \\ 
NGC 6758    &      56     &        104     &      93      &  70  \\  
NGC 6776    &      68     &        107     &      77      &  <30 \\ 
NGC 6851    &      90     &        148     &     126      &  <38 \\ 
NGC 6868    &     141     &        190     &     154      &  <32 \\ 
NGC 6876    &     125     &        192     &     107      &  --  \\ 
NGC 7041    &     135     &        306     &     276      &  <31 \\ 
NGC 7562    &     107     &        131     &     112      &  <31 \\  
NGC 7619    &     102     &        196     &      94      &  <33 \\ 
NGC 7626    &     110     &        235     &     161      &  <49 \\ 
NGC 7796    &     108     &        175     &      86      &  <19 \\ 
IC 4889     &      70     &        121     &    125       &  <30 \\ 
IC 5105     &      66     &         88     &     98       &  <26 \\ 
\hline
\end{tabular}
\caption{ Total integrated fluxes  inside the field of view for the galaxies in
  the sample.}
\label{table:fluxos}
\end{center}
\end{table}
\end{footnotesize}

\section{Spectral energy distributions}
\label{sec:sed}

\begin{figure*}
  \resizebox{\hsize}{!}{\includegraphics[width=\textwidth]{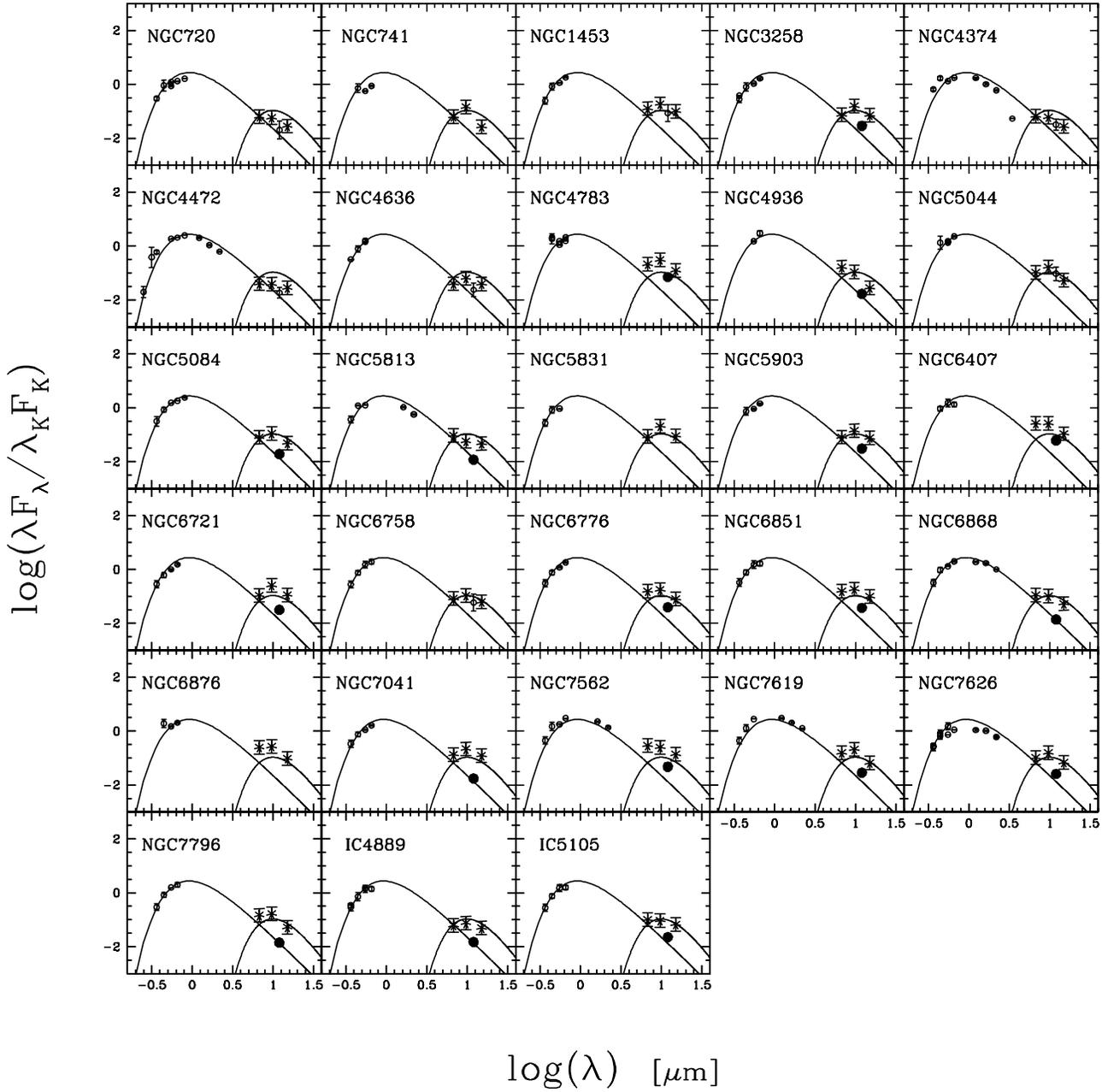}}
\caption{ Spectral energy distribution of the early-type galaxies in the
  sample. The BB corresponding to the mid-IR has a single temperature (260 K)
  for all galaxies.}
\label{fig:sedinv}
\end{figure*}

\begin{figure}
  \resizebox{\hsize}{!}{\includegraphics{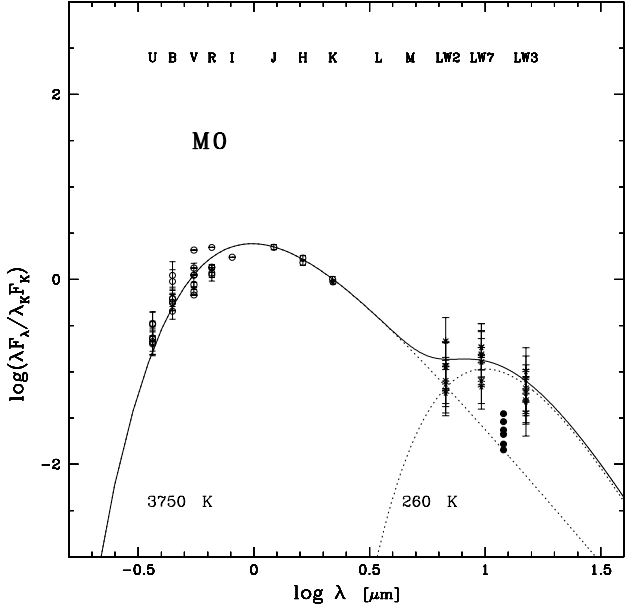}}
\caption{SED for the group M0 (BB with $T=3750$ K) and mid-IR component with a 
  $T=260$ K composite BB. Dots are NED and stars are ISO data. The black dots
  are IRAS ($12\mu{}$m) upper limits.}
\label{fig:sed1}
\end{figure}

\begin{figure}
  \resizebox{\hsize}{!}{\includegraphics{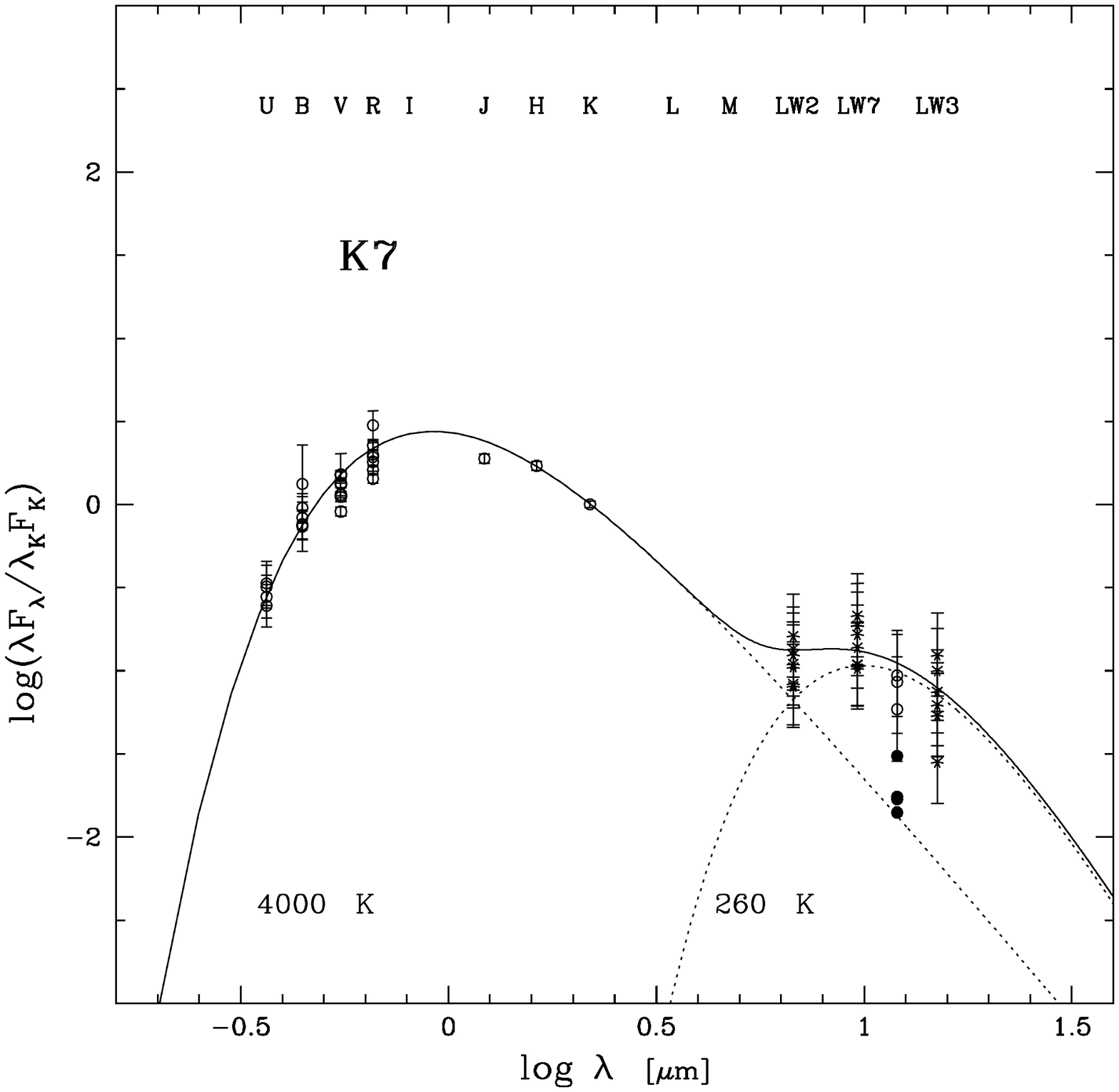}}
\caption{Same as Fig \ref{fig:sed1} for the  K7 galaxy group.}
\label{fig:sed2}
\end{figure}

\begin{figure}
  \resizebox{\hsize}{!}{\includegraphics{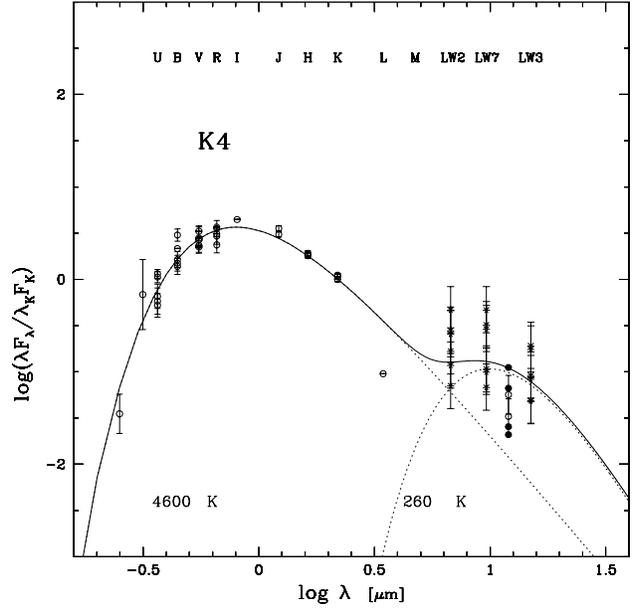}}
\caption{Same as Fig \ref{fig:sed1} for the  K4 galaxy group.}
\label{fig:sed3}
\end{figure}

In order to estimate the relative contributions of the stellar component and
dust emission to the mid-IR luminosity in early-type galaxies, we have built the
spectral energy distribution (SED) for each galaxy using our ISOCAM data
together with UV, optical, near and IRAS IR data taken from the NASA
Extragalatic Database\footnote{The NASA/IPAC Extragalactic Database (NED) is
  operated by the Jet Propulsion Laboratory, California Institute of Technology,
  under contract with the National Aeronautics and Space Administration}. We
have taken from NED the optical and near-IR magnitudes integrated inside an
aperture similar to that corresponding to the ISOCAM field of view
($108^{\prime\prime}$).  UV and IRAS data are total integrated values. Some IRAS
values (black dots in Figures \ref{fig:sedinv}, \ref{fig:sed1}, \ref{fig:sed2}
and \ref{fig:sed3}) are upper limits.  All fluxes were normalized to the near-IR
K magnitude. We then superposed black body (BB) curves to each galaxy SED and
found that the UV, optical and near-IR data for all galaxies can be well matched
with BBs of three different temperatures T=3750 K, T=4000 K and T=4600 K. These
temperatures correspond to dominant stellar populations of spectral types M0, K7
and K4, respectively.  The mid-IR data have been fitted ($\chi^2$) with a
composite BB curve $B_{\nu}(T)\,Q_{\nu} = B_{\nu}(T)\,\nu^{1.6}$ for isothermal
grains (Natta \& Panagia 1976) with T=260 K, for which the hot dust is an
inportant contributor.  For more than $30\% $ of the galaxies in the sample
there are no published near-IR observations, therefore to normalize the SED to
the K band, we have estimated the K flux from the BB curve which fits the
observed UV and optical data.  This normalization was not applied to the BB with
$T=260$ K because at this temperature the K flux is too small and the
normalization would produce high mid-IR values for the BB. Besides, we are
interested in the color temperature rather than in absolute fluxes.  The SED for
each galaxy is shown in Fig.  \ref{fig:sedinv} along with the corresponding BB
curves for the stellar population and dust. There is a remarkable uniformity
among the galaxies with respect to the dust emission properties, except for NGC
720, NGC 4374, NGC 4472 and NGC 4636 whose sizes are much greater than ISOCAM
FOV, leading to underestimated fluxes.  We have arranged the galaxy SEDs in
three different groups named M0, K7 and K4, which are shown in Figures
\ref{fig:sed1} to \ref{fig:sed3}, according to the effective temperature of the
stellar population that represents its SED.  Together with the data points we
have plotted black body radiation curves to illustrate the stellar population
contribution in comparison with the dust component.

It is interesting to stress that all SEDs of our elliptical galaxies are fitted
from the UV to the mid-IR with only two components, one representing the
dominant stellar population continuum and the other the hot dust emission.  The
dust component is remarkably uniform in temperature for all the sample, $T\simeq
260$ K.  We will therefore use this temperature to estimate the mass of the hot
dust (Section 7). In our Galaxy, the far-IR emission comes from dust heated by
the interestellar radiation field (ISRF) (T$<$22 K) and from dust associated
with O and B stars and molecular clouds ($22 {\rm K} < T < 29 {\rm K}$)
(Sodroski et al.  1997).  In order to estimate the pure dust emission, we have
subtracted, from the observed mid-IR fluxes, the underlying stellar
contribution, represented by the corresponding black body curve for that galaxy
group.

The typical mean stellar contributions relative to the total mid-IR fluxes
($F^\star_\lambda + F^{\rm dust}_\lambda $) among the galaxy groups are about
58\% at 6.75 $\mu{}$m, 22\% at 9.63 $\mu{}$m at and 10\% 15 $\mu{}$m.  The infrared emission
of our sample galaxies shows an appreciable contribution of the dust, a
different behavior compared with the early-type galaxies in the sample of
Boselli et al. (1998) or Madden et al.  (1999), for example. This can be
attributed to the fact that our sample was previously selected to be composed of
dust-rich galaxies, therefore their emission in the IR is expected not to be
purely stellar.  The figures presented here use the
corrected flux values.

\section{The mid-IR luminosity distribution }
\label{sec:lumin}

The total mid-IR luminosities (stars+dust emission) have been computed from the
integrated fluxes according to:
\begin{equation}
L_{\nu} = 4\pi D^{2} \, F_{\nu} \, \delta_{\nu} 
\end{equation}
Here $\delta_{\nu}$ is the filter width as follows: $\delta_{6.75\mu{}{\rm
    m}}=16.18$ THz, $\delta_{9.63\mu{} {\rm m}}=6.61$ THz and \mbox{$\delta_{15\mu{}
    {\rm m}}=6.75$ THz}. The fluxes $F_{\nu}$ have been integrated with an
aperture corresponding to the field of view width ($108^{\prime\prime}$).  The
luminosities are listed in Table \ref{table:lumin}, where the total mid-IR luminosity
is $L_{\rm MIR} = L_{6.75} + L_{9.63} + L_{15}$. The observed mid-IR luminosity
ranges from $3-42 \times{} 10^8$ $L_\odot$, which is similar to what Boselli et al.
(1998) found for 6 Virgo Cluster early-type galaxies.

\begin{footnotesize}
\begin{table}
\begin{center}
\begin{tabular}{l  >{$}r<{$} >{$}r<{$} >{$}r<{$}  >{$}r<{$}  } \hline
Galaxy    & L_{6.75\mu{}   m} & L_{9.63\mu{}   m}  &     L_{15\mu{}   m}  & L_{\rm MIR} \\ 
      &  10^8 \; L_\odot  &  10^8 \; L_\odot  &  10^8 \; L_\odot  &  10^8 \;  L_\odot \\ \hline
%
NGC720  &  2.1 &  2.1 &  1.9 &  6.2 \\ 
NGC741  &  9.7 &  23.8 &  8.2 &  41.7 \\ 
NGC1453  &  6.7 &  10.9 &  10.7 &  28.3 \\ 
NGC3258  &  2.0 &  4.6 &  3.9 &  10.5 \\ 
NGC4374  &  1.0 &  1.0 &  0.8 &  2.7 \\ 
NGC4472  &  1.7 &  1.7 &  2.3 &  5.7 \\ 
NGC4636  &  1.3 &  2.3 &  2.5 &  6.1 \\ 
NGC4783  &  7.3 &  11.5 &  8.5 &  27.4 \\ 
NGC4936  &  6.4 &  4.7 &  2.2 &  13.4 \\ 
NGC5044  &  3.9 &  6.7 &  4.0 &  14.6 \\ 
NGC5084  &  1.7 &  2.5 &  2.1 &  6.3 \\ 
NGC5813  &  2.5 &  1.6 &  2.5 &  6.6 \\ 
NGC5831  &  1.2 &  3.1 &  2.5 &  6.8 \\ 
NGC5903  &  2.4 &  4.4 &  4.4 &  11.2 \\ 
NGC6407  &  8.8 &  9.5 &  7.2 &  25.5 \\ 
NGC6721  &  4.5 &  11.2 &  9.2 &  24.9 \\ 
NGC6758  &  2.9 &  4.1 &  4.4 &  11.5 \\ 
NGC6776  &  9.2 &  11.0 &  9.3 &  29.5 \\ 
NGC6851  &  3.2 &  4.0 &  4.0 &  11.3 \\ 
NGC6868  &  5.1 &  5.2 &  4.9 &  15.2 \\ 
NGC6876  &  8.8 &  10.2 &  6.7 &  25.7 \\ 
NGC7041  &  2.2 &  3.8 &  4.0 &  10.0 \\ 
NGC7562  &  6.1 &  5.7 &  5.7 &  17.5 \\ 
NGC7619  &  5.8 &  8.5 &  4.8 &  19.1 \\ 
NGC7626  &  6.3 &  10.2 &  8.2 &  24.7 \\ 
NGC7796  &  5.2 &  6.4 &  3.7 &  15.4 \\ 
IC4889  &  2.0 &  2.6 &  3.2 &  7.9 \\ 
IC5105  &  8.8 &  8.9 &  11.6 &  29.3 \\ 
\hline
\end{tabular}
\caption{ Partial and integrated total luminosities corrected by stellar
  contribution.}
\label{table:lumin}
\end{center}
\end{table}
\end{footnotesize}

\begin{figure}
 \resizebox{\hsize}{!}{\includegraphics{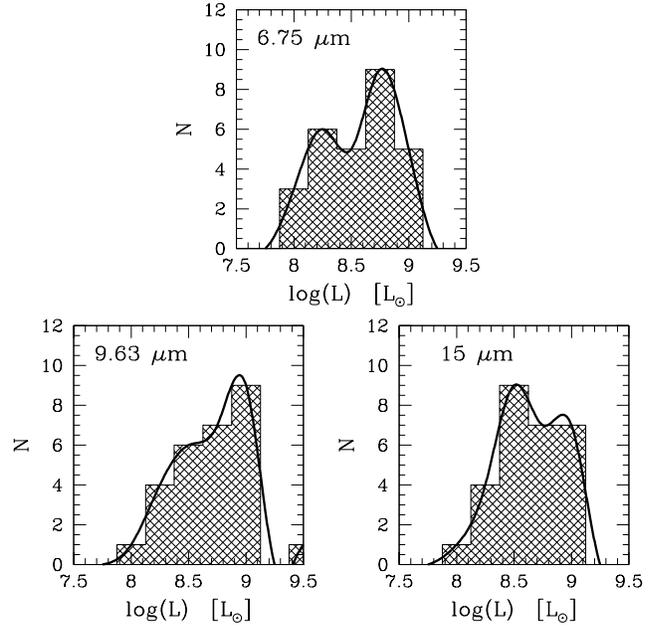}}
\caption{ Luminosity histograms in the three filters for sample  galaxies. 
The solid lines correspond to a spline fit to the points.}
\label{fig:lirhist}
\end{figure}

\begin{figure}
 \centering \resizebox{\hsize}{!}{\includegraphics{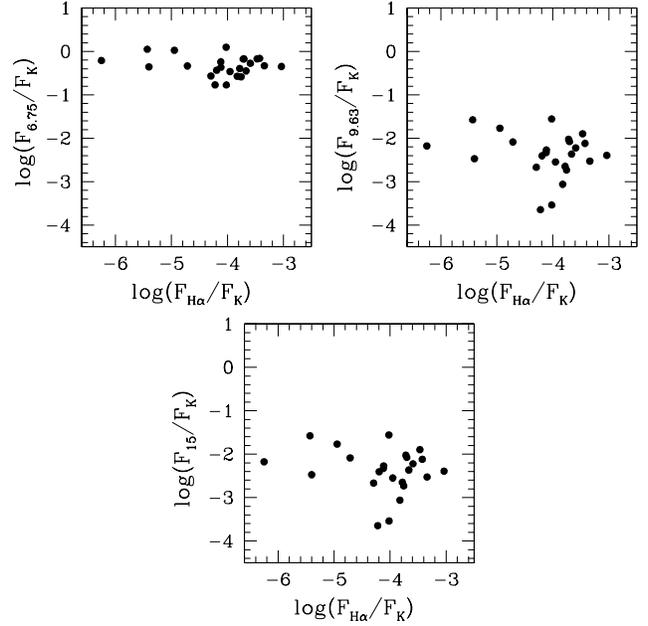}}
\caption{Correlation between the mid-IR and the H$\alpha$ luminosities.
  The vertical axis is the ratio between both luminosities to remove any
  distance effects.}
\label{fig:lirlha}
\end{figure}

Figure \ref{fig:lirhist} shows the histograms of the luminosity distribution for
the observed galaxies. The plots shows the luminosities that have been corrected
by subtracting the stellar component.  It can be seen that the three
luminosities have a similar mean value $\log\langle L \rangle \simeq 8.6
\;L_{\odot}$. The source that dominates the SED at 6.75 $\mu{} $m is probably the
Rayleigh-Jeans tail of the old stellar population. For 15 $\mu{} $m the well defined
peak at $\log\langle L \rangle \simeq 8.5$ may indicate that for most of the
galaxies in our sample the dust properties such as temperature mass and
composition are similar.  In order to see whether the ionizing source of the gas
is also responsible for heating the dust, we studied the correlation between the
mid-IR fluxes corrected by the stellar contribution and H$\alpha$ fluxes.
Figure \ref{fig:lirlha} shows such a flux-flux plot, normalized by the K flux.
Normalized $6.75\mu{}$m fluxes are constant for 3 decades of H$\alpha$. At $9\mu{}$m and
$15\mu{}$m the weakest H$\alpha$ do not correlate, although for galaxies in the
range $-4.5 < \log(F_{{\rm H}\alpha} / F_{\rm K})<-3.5$ there is a weak
correlation. At least for some galaxies the source of gas ionization and dust
heating must be the same and also that both gas and dust coexist in the same
regions.

\begin{figure}
  \centering \resizebox{\hsize}{!}{\includegraphics{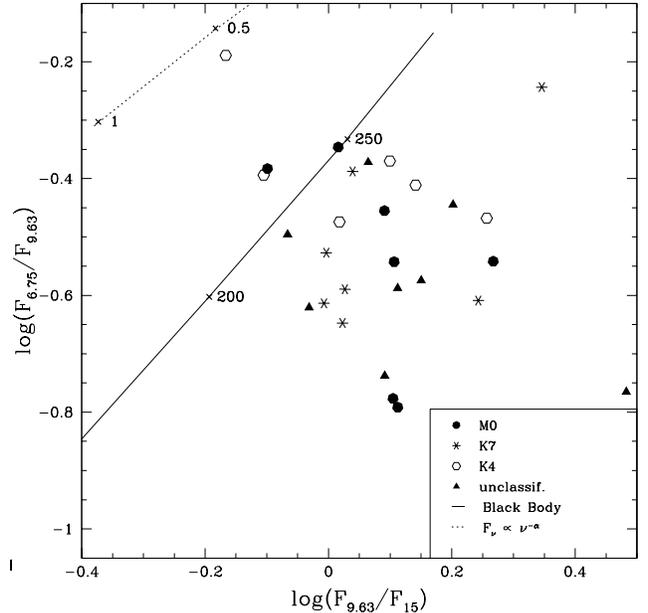}}
\caption{Color-Color diagram for the sample galaxies. 
  the line corresponds to BB at different temperatures. See text for description
  of the models}
\label{fig:corcor}
\end{figure}

\section{Mid-IR colors}
\label{sec:midir}

The integrated properties of mid-IR dust emission in galaxies can be analysed by
their colors, after subtracting the stellar population contribution. Figure
\ref{fig:corcor} illustrates the color-color diagram for the observed 6.75, 9.63
and 15 $\mu{} $m fluxes, along with he corresponding colors of pure black body
emission for different temperatures and power-law models. Clearly, our observed
$F_{9.63}/F_{15}$ colors are bluer than those expected for a pure black body
emission, concentrated near the BB with $T=260$ K. Galaxies belonging to
different stellar population groups (M0, K7, K4) can be found in the same
regions of the color-color diagram. Since these colors have been obtained after
subtracting the stellar population contribution they are associated to
properties of the dust, such as chemical composition, optical properties, grain
size and the nature of the heating source.

The blue excess of the mid-IR colors (in particular the $F_{9.63}/F_{15}$ ratio)
of our sample of galaxies with respect to the pure black body curve can be
explained if we assume that the emission is produced by dust composed of
graphite, silicates and PAHs, as modeled by Desert et al. (1990). This model
includes three components: large silicate grains, very small graphite grains
(3-dimensional) and PAHs (2-dimensional). 
There is a silicate band at $9.7\mu{}$m, while the main PAHs features in
this spectral region are at 3.3, 6.2, 7.7, 8.6 and 11.3 $\mu{} $m (Puget, Leger and
Boulanger 1985), which are covered by the filters used in the present
observations.  The scatter of points, for different galaxies, may be attributed
to differences in the amount and/or distribution of the radiation responsible
for heating the dust and PAHs. As a result, the contribution of molecules to the
mid-IR emisson will vary. The remaining scatter in the correlation can be easily
accounted for by differences in the interestellar radiation field and dust
properties.

\section{Dust mass estimates}
\label{sec:mass}

In the presence of a radiation field, dust grains absorb and scatter varying
fractions of the incident photons, thus changing the intensity and spectral
character of the radiation spectrum. Part of the absorbed energy is re-emitted
in the infrared. Consequently, an analysis of the emission and absorption
properties of dust grains can be used to derive estimates on the amount of dust
present in galaxies.

This dust can be found in several components (Section \ref{sec:midir}), e.g.
graphites, silicates, PAHs, UIBs. However, our present photometric data (broad
band fluxes) do not allow us to distinguish among these components, for which it
would be necessary high spectral resolution spectroscopy.  Besides, recent
published ISO observations and models by Athey et al. (2002) indicate that
amorphous silicate are the major constituent of dust in the oxygen-rich
environments in the envelopes of low mass AGB stars of early-type galaxies,
where the dust is. They also have found no evidence of PAH emission in the
strongest band 7.7$\mu{}$m with an detection limit of 5 mJy. Accordingly, we
estimate the dust mass using silicate grains as a probe; grains have similar
properties, so the dust mass would not be appreciably different if we had 
considered graphite grains.  The morphology of the ionized gas, the hot and cold
dust as well as the luminosity distribution will be analyzed in detail in a
paper in preparation.

A typical large dust grain absorbs photons of wavelength $\lambda\leq 1 \;\mu{} $m,
heats to a certain temperature and re-emits energy in the infrared approximately
as a black-body. Our sample galaxies present a maximum in their SEDs in the
mid-IR at $\lambda\approx10.5\,\mu{} $m (Fig. \ref{fig:sedinv}), which corresponds
to a grain temperature $T=260$ K. In the context of this paper, grains heated to
that temperature are considered as hot dust. Thus, if $L^{\rm g}_{\rm MIR}$\ is
an individual grain's MIR luminosity and $L_{\rm MIR}$\ the integrated
luminosity of a given galaxy observed in the range 5--18 $\mu{} $m\ (Table
\ref{table:lumin}), the total number of emitting grains (hot dust) can be
approximated as
\begin{equation}
  N_{HD} \approx \frac{ L_{\rm MIR} }{ L^{\rm g}_{\rm MIR} }.
\end{equation}

A single grain luminosity is calculated using a standard emission model
(Panagia 1975, Bollea and Cavaliere 1976,  Barvainis 1987) in which
\begin{equation}
  L^{\rm g}_{\rm MIR} = \int_{\nu_{\rm min}}^{\nu_{\rm max}} 4 \pi a^2 
  Q_{\nu}\pi B_{\nu}(T)\,d\nu, \label{eq:LgMIR}
\end{equation}  
where the absorption efficiency in the infrared can be approximated by (Draine and
Lee 1984)
\begin{equation}
  Q_{\rm abs}  =  3 \; \frac{a }{\lambda^{1.6}} 
\end{equation}
where $a$ and $\lambda$ are given in $\mu{}$m. $B_{\nu}(T)$ is Planck's black-body
function and $a$\ is the grain radius.  Finally, for grains with density
$\rho_{\rm g}$, the hot dust mass is
\begin{eqnarray}
  M_{HD} \approx \frac{4 \pi}{3} a^3\;N_{HD}\; \rho_{\rm g}.  
\end{eqnarray}

Further assumptions concerning the grain properties are necessary in order to
estimate the mass are (Ferrari et al. 1999 and references therein): a) the
grains are assumed to be composed mainly of silicates for which $\rho_{\rm g} =
2550 {\rm \;kg\;m}^{-3}$; b) a single grain size is considered as a first
approximation, $a = 0.00196 \; \mu{} $m.  In fact, this is the median size $\langle
a \rangle$ of a distribution \mbox{$f(a)\propto a^{-7/2}$} with
$0.0012<a<0.015\,\mu{} $m; c) a single grain temperature $T=260$ K -- corresponding
to the hot dust component -- is used and d) equation (\ref{eq:LgMIR}) is
integrated between $5 < \lambda < 18 \;\mu{} $m, since this is the range covered by
the ISO filters used.

The resulting dust masses for each galaxy are shown in
Table~\ref{table:mirmass}.  We note that the masses of the hot dust component
($T=260$ K) range between $\sim 10$\ and $\sim 200\ {\rm M_\odot}$.

\begin{table}
\caption[]{Dust mass}
\begin{center}
\renewcommand{\tabcolsep}{2.7mm}
\begin{tabular}{lcccc} 
\hline\hline
Galaxy & R   & $M_{HD}$  & $M_D$& $M_{A_V}$ \\
\cline{2-5}&  \\
&(kpc)& $(M_\odot)$ &$(M_\odot)$ &$(10^3\,M_\odot)$ \\
\hline
NGC 720  &  3.3   &   28   &   36    &   ---      \\ 
NGC 741  &  11.0  &   192  &  395    &   ---      \\ 
NGC 1453  &  7.8  &   130  &  204    &   ---      \\ 
NGC 3258  &  5.5  &   48   &  100    &   ---      \\ 
NGC 4374  &  2.0  &   12   &   10    &   27$^\dagger$    \\ 
NGC 4472  &  2.4  &   26   &   14    &   1        \\ 
NGC 4636  &  2.4  &   28   &   18    &   ---      \\ 
NGC 4783  &  9.2  &   126  &  276    &   ---      \\ 
NGC 4936  &  6.6  &   62   &  148    &   ---      \\ 
NGC 5044  &  5.8  &   67   &  114    &   12       \\ 
NGC 5084  &  3.4  &   29   &   45    &   ---      \\ 
NGC 5813  &  3.9  &   30   &   37    &   13       \\ 
NGC 5831  &  3.9  &   31   &   57    &   ---      \\ 
NGC 5903  &  4.9  &   52   &   81    &   15       \\ 
NGC 6407  &  8.9  &   117  &  200    &   ---      \\ 
NGC 6721  &  8.5  &   114  &  279    &   ---      \\ 
NGC 6758  &  6.5  &   53   &  144    &   4        \\ 
NGC 6776  &  10.5 &   136  &  422    &   ---      \\ 
NGC 6851  &  5.4  &   52   &   73    &   ---      \\ 
NGC 6868  &  5.4  &   70   &   98    &   ---      \\ 
NGC 6876  &  7.5  &   118  &  185    &   ---      \\ 
NGC 7041  &  3.6  &   46   &   44    &   ---      \\ 
NGC 7562  &  6.8  &   80   &  178    &   ---      \\ 
NGC 7619  &  6.8  &   88   &  178    &   ---      \\ 
NGC 7626  &  6.8  &   114  &  150    &   ---      \\ 
NGC 7796  &  6.3  &   71   &   98    &   ---      \\ 
 IC 4889  &  4.8  &   36   &   75    &   35       \\ 
 IC 5105  & 10.4 &   135   &  414    &   130      \\ 
\hline
\end{tabular}
\begin{list}{Table Notes.}
\item{ Dust masses derived from the grain luminosity ($M_{HD}$), effective 
extinction ($M_D$) and optical extinction ($M_{A_V}$). R is the radius of the 
MIR emitting region.  \\ $\dagger$ Goudfrooij et al. 1994a}
\end{list}
\label{table:mirmass}
\end{center}
\end{table}

Assuming now another picture in which stars and dust are uniformly
distributed throughout the galaxy, the effective extinction produced by the dust
distribution on the stellar radiation field can be used to estimate the dust
mass. According to this scenario, the dust luminosity ($L_{D}$) corresponds to
the absorbed fraction of the intrinsic galaxy luminosity ($L^*_{\rm gal}$, assumed
to come primarily from the stars), as follows:

\begin{equation}
\frac{L_{D}}{L^*_{\rm gal}} = 1-\frac{1-e^{-\tau^\prime}}{\tau^\prime},
\end{equation}
\begin{equation}
\tau^\prime=\frac{4}{3}\pi a^2 \; n_{\rm g} \; Q_e  \; R
\end{equation}
where $\tau^\prime$ is the effective optical depth, $n_{\rm g}$\ is the grain
number density, $Q_e$\ is the effective absorption efficiency and $R$\ is the
radius of the emitting region. We consider as a first approximation $L^*_{\rm
  gal} = L_{\rm gal} + L_D$. Both $L_{\rm gal}$\ and $L_D$\ are calculated
integrating the corresponding curves (Fig. \ref{fig:sed1}-\ref{fig:sed3}) along
the entire wavelength range. $\tau^{\prime}$ values are, respectively, 0.058,
0.051, 0.038 and 0.049 for the M0, K7, K4 and unclassified groups. Accordingly, the
dust mass is calculated as follows:

\begin{equation}
M_D = \frac{4}{3}\pi R^3 \,  n_{\rm g} \, \frac{4}{3}\pi a^3\rho_{\rm g} 
= \frac{4}{3} \pi a \; \frac{\tau^\prime}{Q_e} \;  R^2  \rho_{\rm g}.
\end{equation}

The radius of the emitting regions are measured on the ISO images and are given
in the second column of Table~\ref{table:mirmass}; grain parameters are those of
silicates with $a=0.00196 \mu{}$m\ and $\rho_{\rm g}=2550{\rm \;kg\;m}^{-3}$. The
resulting dust masses are given in Table~\ref{table:mirmass}.  There is a
good agreement between the dust mass values obtained with both methods.

Seven galaxies in the present sample are in common with the sample studied in
Paper\,II and one with the sample of Goudfrooij et al. (1994a). For galaxies in
Paper II the dust masses have been estimated from the optical extinction (which
is caused mostly by cold dust $T\sim 30$ K composed of big silicate grains with
$a_{\rm si}=0.1 \mu{} $m).  These values ($M_{A_V}$) are shown, for comparison
purposes, in the last column of Table~\ref{table:mirmass}. Dust masses estimated
from the effective extinction are comparable to those corresponding to the hot
dust, and 100 times in average smaller than those derived from the optical
extinction.




We conclude that the hot dust mass component in early-type galaxies corresponds
to a small fraction (a few percent) of the cold dust component. Although this is
a new result for elliptical galaxies, it is not surprising since a similar
effect has been observed in spiral galaxies such as M\,31, in which the warm
dust mass, derived from the 12 and 25 $\mu{} $m IRAS fluxes, is a factor $\sim 240$
smaller than the cold dust component derived from 60 to 175 $\mu{} $m fluxes (Haas
et al. 1998).

\section{Conclusions}
\label{sec:conclusions}

We have shown the results of the mid-IR observation of 28 luminous early-type
galaxies directed to study the properties of their interstellar medium. The
sample spans a large interval in optical luminosity and includes both radio-loud
and radio-quiet, X-ray emitters and non-emitters. From our observations, which
consist of 6.75 $\mu{} $m, 9.63 $\mu{} $m and 15 $\mu{} $m images, we have mapped the mid-IR
dust emission and luminosities.  Most of the galaxies in our sample contain
extended and measurable amounts of dust in the three mid-IR wavelengths. We have
built the spectral energy distribution for each galaxy from the UV to the
mid-IR, and found that the SED's in the wavelength interval from the UV to the
near-IR can be well matched by BB curves of temperatures T=3750, T=4000 and
T=4600 K.  These temperatures correspond respectively to dominant stellar
populations of spectral types M0, K7 and K4.  The mid-IR data have been fitted
with a composed BB with T=260 K, for which the hot dust is a significant
contributor.  This curves is a reasonable fit for all the galaxies in the
sample.  We have estimated that the mean stellar contribution to the total
mid-IR flux is about 58 \% at 6.75 $\mu{} $m , 22 \% at 9.63 $\mu{} $m and 10 \% at 15
$\mu{} $m.  The observed mid-IR luminosity ranges from $3 - 42 \times{} 10^{8}$ $L_\odot$.
We compared the mid-IR colors with the BB colors for different temperatures and
showed that our observed $F_{9.63}/F_{15}$ colors are bluer than those expected
for a pure black body emission. The mid-IR colors of our sample of galaxies can
be explained if we assume that the emission is the result of dust composed of
graphite, silicates and PAHs as modeled by Desert et al. (1990). The dust mass
of the mid-IR component was estimated assuming that the mid-IR
emission comes basically from silicates  grains heated at ($T=260$ K). The
mass of the hot dust component estimated in this paper ranges between 10 and 200
$M_\odot$ and we conclude that it corresponds to a very small fraction (a few
percent) of the cold dust component ($T\simeq 40\,K$) calculated by Merluzzi
(1998) for a sample of E galaxies.  Using the concept of effective optical
depth, we found dust masses in the range $10 - 400\;M_{\odot}$.  This is in very
good agreement with mass calculated from the grain luminosity.  Finally, these
observations have shown that in addition to gas and cold dust, the ISM of
elliptical galaxies also includes a measurable amount of hot dust.

\section{Acknowledgements}
\label{sec:aknowledgements}

We thank M. Sauvage for providing patience assistance with the CIA packages;  D.
Calzetti and S. Arribas for interesting discussions and suggestions during the
preparation of this work.  This research has
been partial supported by PRONEX/FINEP/CNPq grant 7697100300 and by the STScI
visitor program.


\end{document}